\documentclass[conference,table]{IEEEtran}
\usepackage[utf8]{inputenc}
\usepackage{textgreek}
\IEEEoverridecommandlockouts
\usepackage{bbding}
\usepackage{makecell}
\usepackage{cite}
\usepackage{amsmath,amssymb,amsfonts}
\usepackage{bm}
\usepackage{textcomp}
\usepackage{graphicx}
\usepackage{pgfplots}
\pgfplotsset{compat=newest}
\usepackage{subcaption}
\usepackage{multirow}
\usepackage{adjustbox}
\usepackage{booktabs}
\usepackage{siunitx}
\usepackage{tabularx}
\usepackage{comment}
\usepackage{authblk}
\usepackage{tikz}
\usetikzlibrary{arrows.meta, positioning, shadows}
\usepackage[skip=3pt]{caption}
\usepackage{hyperref}
\hypersetup{colorlinks=true, citecolor=blue, linkcolor=blue, urlcolor=blue}
\usepackage{float}

\usepackage{fancyhdr}

\fancypagestyle{firstpage}
{
    \fancyhead[L]{\footnotesize Personal use of this material is permitted.  Permission must be obtained for all other uses, in any current or future media, including reprinting/republishing this material for advertising or promotional purposes, creating new collective works, for resale or redistribution to servers or lists, or reuse of any copyrighted component of this work in other works.This manuscript has been submitted to an IEEE Transactions journal and is currently under review.}
    \fancyhead[R]{}
}
\begin{document}

\title{WARD: Runtime \underline{W}orkload-\underline{A}daptive Vision T\underline{R}ansformer Framework for \underline{D}ependable Edge AI}
\author{Mahdi~Taheri, Pramit Kumar Bhaduri, Mohammad Masoumi, Ali Mahani%
\thanks{Mahdi Taheri  is with Humboldt University of Berlin and also with Tallinn University of Technology, Tallinn, Estonia. Pramit Kumar Bhaduri is with Brandenburgische Technische Universität Cottbus-Senftenberg, Germany. Mohammad Masoumi and Ali Mahani are with Shahid Bahonar University of Kerman, Iran. (e-mail: mahdi.taheri@taltech.ee; bhadupra@b-tu.de; m.masoumi@eng.uk.ac.ir; mahani@eng.uk.ac.ir).}}



\maketitle
\thispagestyle{firstpage}

\begin{abstract}

Edge-deployed AI operate under dynamically changing power budgets, reliability requirements, and input distributions, requiring continuous adaptation. Such conditions arise in long-running edge AI applications, including autonomous systems, industrial monitoring, and satellite onboard intelligence. Existing fault-tolerant methods assume static operating conditions, whereas continual learning techniques neglect concurrent hardware faults during online adaptation. Moreover, the practical deployment of runtime-adaptive reliability frameworks on programmable AI accelerators remains largely unexplored.

 This paper presents \textit{WARD}, a runtime-adaptive Vision Transformer framework that combines channel-wise subnetwork partitioning, reliability-aware continual learning, and dynamic operating-mode scheduling to jointly optimize performance, fault tolerance, and adaptation according to runtime conditions. Two physically isolated subnetworks execute under four operating modes (i.e. Full-Precision Mode, Low-Power Mode, High-Reliability Mode, and Adaptive Mode) that dynamically adjust computational cost and reliability while ensuring uninterrupted inference for real-time requirements. To validate the practical deployability of the proposed framework, WARD is implemented on a lightweight FPGA-based accelerator extended with runtime hardware support for mode scheduling and resource management. Experimental results demonstrate that the proposed split architecture achieves a network-level failure rate of only 1.79\% under high Bit Error Rates. The hardware implementation incurs less than 5\% area overhead and supports runtime mode transitions within few clock cycles, demonstrating that adaptive reliability management can be integrated into programmable edge AI accelerators with negligible implementation overhead.

\end{abstract}


\section{Introduction}

Vision Transformers (ViTs) \cite{dosovitskiy2021vit} have become one of the dominant architectures for image recognition by extending the self-attention mechanism originally developed for sequence modelling to image patches \cite{vaswani2017attention}. Their ability to capture long-range spatial dependencies enables state-of-the-art performance across numerous vision applications, including autonomous driving, medical image analysis, robotic perception, industrial inspection, and satellite-based Earth observation and autonomous navigation \cite{furano2020edge}. As these applications move from cloud infrastructures to resource-constrained edge platforms, the deployment objective extends beyond inference accuracy alone. Edge AI systems operate continuously under strict power, latency, and reliability constraints while interacting with dynamic physical environments. In many of these applications, inference outputs directly drive safety-critical or mission-critical decisions, making uninterrupted operation equally as important as prediction accuracy. Consequently, both the neural network and its execution platform require the ability to adapt their behaviour during operation without interrupting inference.

\begin{figure*}[t]
    \centering
    \includegraphics[width=2\columnwidth]{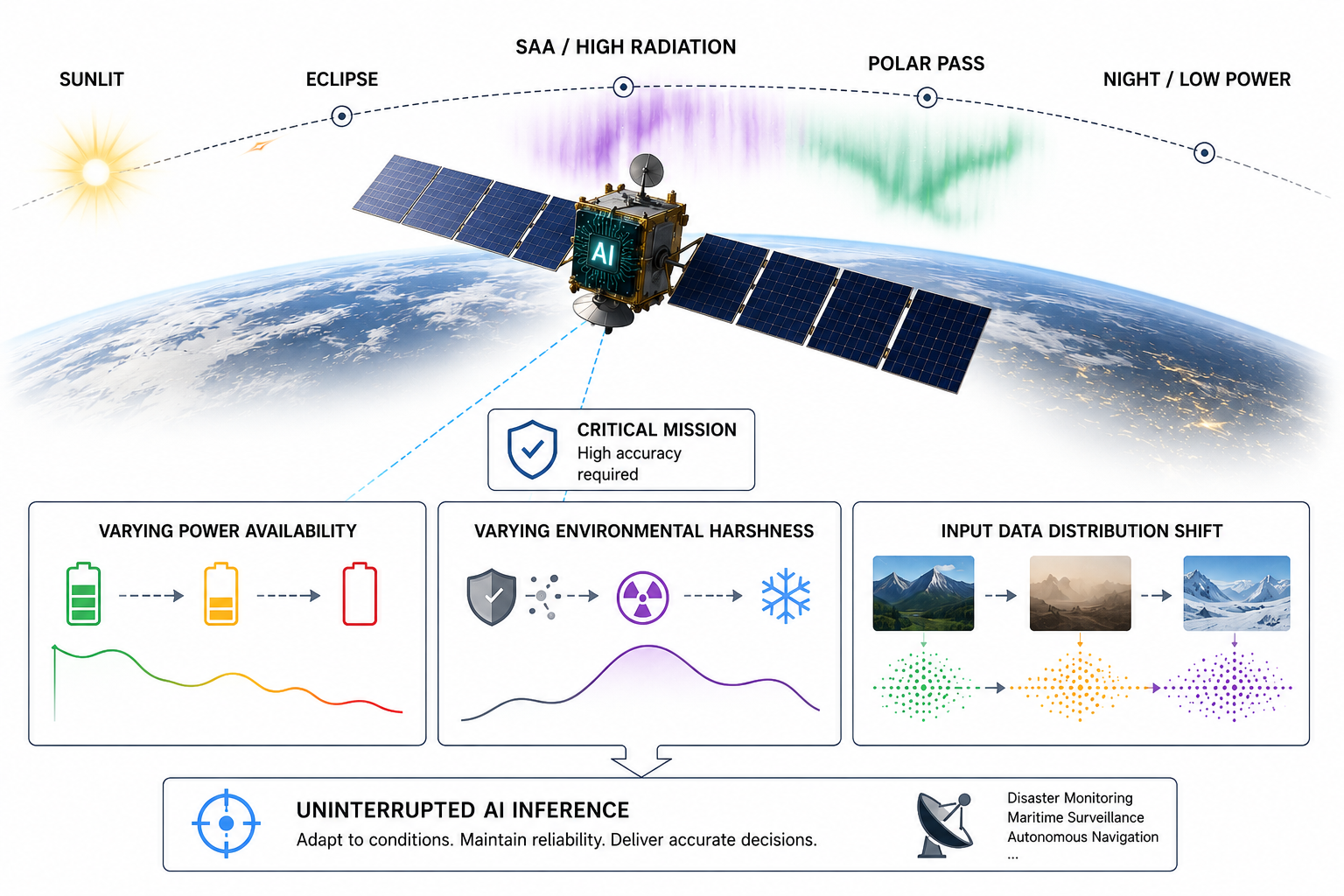}
    \caption{Dynamic operating conditions of a mission-critical edge AI application.}
    \label{fig:motivation}
\end{figure*}

Satellite edge intelligence represents one of the most demanding deployment scenarios because it combines these requirements within a single application (Fig. \ref{fig:motivation}). Commercial off-the-shelf (COTS) processors enable onboard neural network inference at substantially lower cost and higher computational capability than radiation-hardened processors \cite{denby2020orbital,diana2024review}. This capability enables real-time onboard processing for applications such as disaster monitoring, wildfire detection, maritime surveillance, and autonomous satellite navigation without relying on delayed ground-station communication \cite{mateo2023orbit}. However, a satellite continuously traverses orbital regions with different power availability, radiation intensity, and environmental conditions. Solar power disappears during eclipse, radiation intensity varies significantly across orbital regions, and the observed data distribution changes with cycle variations, illumination conditions, and sensor aging. Unlike conventional edge systems, these changing conditions occur without the possibility of manual intervention, making continuous and uninterrupted inference a fundamental operational requirement rather than a performance objective. Consequently, the inference framework requires the ability to dynamically adapt its computational behaviour and reliability strategy throughout the mission lifetime while remaining deployable on lightweight edge AI hardware.

Operating under these conditions introduces a persistent hardware reliability challenge, like high-energy particles that induce single-event upsets (SEUs) in processor memories, and corrupting neural network parameters stored in on-chip memories during inference \cite{baumann2005radiation}. Consequently, the probability of hardware faults changes continuously rather than remaining constant throughout execution. Applying the same reliability mechanism under all operating conditions, therefore, introduces unnecessary computational and energy overhead during low-risk periods while potentially remaining insufficient during high-radiation intervals.

Fault injection studies on Vision Transformers demonstrate that hardware-induced parameter corruption produces highly non-uniform behaviour \cite{xue2023softerror,he2025fine, SENTRY}. Most bit-level faults remain functionally benign, whereas a small subset produces catastrophic failures that propagate NaN or Inf values throughout the remaining inference pipeline \cite{liao2025analyzing,tung2026anatomy}. Unlike moderate accuracy degradation, catastrophic numerical failures cannot be mitigated through confidence estimation or post-processing because the entire computational stream becomes invalid. Conventional fault-tolerance techniques, including Triple Modular Redundancy (TMR), improve robustness under static operating conditions but impose fixed computational overhead or assume a constant reliability requirement throughout execution \cite{tedeschi2025safeneureka,shao2024spaceborne,kodamanchili2025adaptive}. None of these approaches adapts the protection strategy according to the dynamically changing reliability requirements encountered during orbital deployment.

A second challenge arises from the need for continuous adaptation throughout the mission lifetime. Satellite sensors continuously observe changing environments under varying seasonal cycles, illumination conditions, orbital geometries, and progressive sensor aging. Consequently, the input distribution gradually diverges from the data available during deployment, causing a progressive degradation in inference accuracy. Continual and incremental learning techniques address this challenge by enabling online parameter adaptation without complete model retraining \cite{kirkpatrick2017overcoming, daniels2023efficient, christophides2024stochastic}. In-orbit demonstrations further confirm that onboard model adaptation is practically achievable \cite{mateo2023orbit}. However, existing continual learning approaches typically assume that adaptation can be performed without compromising the primary inference task \cite{liu2025enabling}, whereas many mission-critical edge AI applications require uninterrupted real-time operation throughout the learning process. Suspending inference while updating model parameters may interrupt safety-critical or mission-critical decision making, rendering conventional adaptation strategies unsuitable for continuous deployment. Furthermore, existing continual learning methods assume reliable hardware execution. When a transient hardware fault corrupts a parameter during adaptation, the resulting error propagates through subsequent gradient updates and permanently contaminates the learned model. Existing continual learning frameworks, therefore, address distribution shift \cite{wang2024comprehensive} while largely neglecting both uninterrupted real-time inference and concurrent hardware faults during online adaptation.

Realising such runtime adaptation also places new requirements on the underlying hardware platform. Existing edge AI accelerators primarily optimise throughput through static execution pipelines and fixed resource allocation \cite{sze2017efficient}, providing limited support for dynamically changing inference modes during execution. Runtime adaptation requires the execution platform to coordinate resource allocation, scheduling, and fault monitoring while maintaining uninterrupted inference. These requirements become particularly challenging in programmable AI accelerators, where multiple processing elements, shared memory hierarchies, runtime scheduling, and resource management introduce complex execution dependencies. Consequently, demonstrating runtime-adaptive inference on an AI accelerator provides strong evidence that the proposed methodology remains deployable while requiring only lightweight architectural support rather than extensive modifications to the computational datapath.

Figure~\ref{fig:motivation} summarizes these challenges. This work introduces \textit{WARD}, a lightweight runtime framework for continuous, fault-aware Vision Transformer inference under dynamic orbital conditions. WARD partitions a pretrained Vision Transformer into two physically isolated subnetworks that operate under four runtime modes, Full-Precision Mode (FP), Low-Power Mode (LP), High-Reliability Mode (HR), and Adaptive Mode (AD), allowing the system to continuously balance computational cost, reliability, and online adaptation according to the current mission conditions. 

To demonstrate that the proposed framework remains practical beyond software simulation, WARD is implemented on a lightweight GPU-like FPGA \cite{kadi2018general} (PERUN) accelerator extended with a minimal runtime support layer for mode scheduling and resource allocation. The implementation validates that the proposed runtime framework integrates into an existing programmable execution platform through lightweight runtime extensions, demonstrating that WARD is readily deployable across a broad range of AI accelerators with minimal architectural modifications.
The main contributions of this work are summarized as follows:

\begin{enumerate}

\item A runtime-adaptive Vision Transformer framework that combines continual learning and dynamic operating-mode scheduling to maintain reliable inference under changing orbital conditions.

\item A channel-wise subnetwork partitioning methodology that provides physical parameter isolation and enables uninterrupted inference through lightweight runtime mode transitions between full precision, low power, high reliability, and adaptive execution.

\item A reliability-aware continual learning strategy that restricts online parameter updates to empirically fault-resilient regions of the parameter space, preserving reliability-critical parameters while enabling continuous adaptation

\item A lightweight hardware implementation of the proposed framework called PERUN, as an extension of a programmable FPGA accelerator (FGPU), demonstrating that runtime reliability management and mode scheduling can be integrated through minimal architectural support while preserving practical deployability on real-world edge AI platforms.

\end{enumerate}

The remainder of this paper is organized as follows. Sections~\ref{sec:background} and~\ref{sec:relatedwork} review the relevant background and related work. Section~\ref{sec:methodology} presents the WARD framework together with its hardware realization. Section~\ref{sec:experiments} evaluates both the runtime framework and its hardware implementation through reliability, adaptation, and deployment experiments. Finally, Section~\ref{sec:conclusion} concludes the paper.


\section{Background}\label{sec:background}

\begin{figure*}[t]
    \centering
    \includegraphics[width=2\columnwidth]{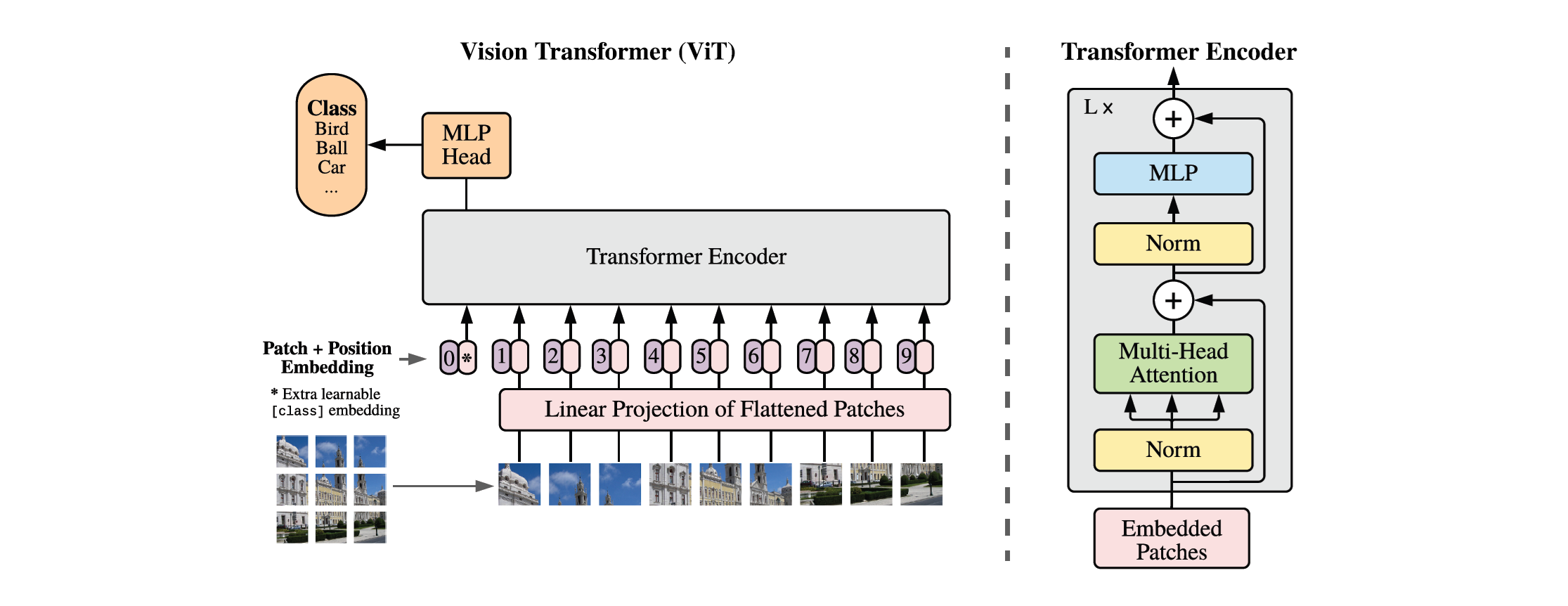}
    \caption{Overview of the Vision Transformer (ViT) architecture, adapted from \cite{dosovitskiy2021vit}.}
    \label{fig:vit_arch}
\end{figure*}

\subsection{Vision Transformer}

The Vision Transformer adapts the encoder architecture introduced for sequence modelling \cite{vaswani2017attention} to image classification by treating an image as an ordered sequence of patch tokens. An input image of dimensions $H \times W \times C$ is divided into $N$ non-overlapping patches of size $P \times P$, which are flattened and linearly projected to a $D$-dimensional embedding space. A learnable classification token (CLS) is prepended to this sequence, a trainable positional encoding is added to each element, and the full sequence passes through a stack of $L$ transformer encoder blocks \cite{dosovitskiy2021vit}. Each block applies two sub-operations in turn. The first is multi-head self-attention (MHSA), which computes query, key, and value projections for all token positions and produces a weighted mixture of value vectors, with weights derived from query-key dot-product similarities. The second is a two-layer multilayer perceptron (MLP) with GELU activations, which applies a position-wise nonlinear transformation to each token independently. Both sub-operations are preceded by Layer Normalization \cite{ba2016layernorm} and are wrapped in additive residual connections. After all $L$ blocks, the final CLS token is passed through a linear classifier head to produce class-level predictions. Figure~\ref{fig:vit_arch} illustrates this structure. While this architecture provides excellent representational capability, its tightly coupled computation also causes all encoder blocks to operate on a shared feature representation, making reliable continuous inference particularly challenging once a hardware fault affects the execution pipeline.

Two structural properties of this design carry direct consequences for hardware fault effects. The residual connections are strictly additive: each encoder sub-block writes its output directly into the main token sequence without any intermediate gating or re-normalization between blocks. A corrupted weight in any attention projection or MLP layer therefore injects an error into the shared representational stream, which then propagates through all subsequent blocks without attenuation. Layer Normalization, applied twice per encoder block at the input to MHSA and MLP and a final time at the network output, normalizes each token's activations across the embedding dimension using per-token statistics and then applies learned per-channel scale ($\gamma$) and shift ($\beta$) parameters. Because these scale and shift parameters are shared across all tokens in the sequence, a fault in any single $\gamma$ or $\beta$ value propagates a consistent rescaling or shift to every token simultaneously, amplifying the fault effect uniformly across the full batch. These structural characteristics concentrate vulnerability at specific architectural locations while tightly coupling the computation across the entire network, making fault isolation, recovery, and uninterrupted inference considerably more difficult than in architectures with naturally separated computational paths.

\subsection{Continual Learning}

Continual learning addresses the problem of updating a trained neural network on new data distributions without erasing the knowledge encoded in its parameters for prior distributions, a failure mode known as catastrophic forgetting \cite{kirkpatrick2017overcoming}. Standard gradient descent is globally destructive: training on a new distribution overwrites parameter configurations that were shaped by earlier data. Three families of methods have been developed to mitigate this. Regularization-based methods add a penalty term to the training loss that resists changes to parameters identified as important for past tasks. Elastic Weight Consolidation (EWC) \cite{kirkpatrick2017overcoming} uses the Fisher information matrix as an importance proxy. Synaptic Intelligence \cite{zenke2017synaptic} accumulates a path-integral-based importance estimate continuously during training. Rehearsal-based methods maintain a buffer of stored exemplars from prior distributions and replay them alongside new training data. iCaRL \cite{rebuffi2017icarl} combines exemplar replay with representation learning to preserve class-discriminative structure across increments. Learning without Forgetting (LwF) \cite{li2018lwf} avoids storing exemplars entirely by using the previous model's soft predictions as supervisory targets. Parameter isolation methods assign separate subsets of the model to separate tasks, preventing interference by construction \cite{christophides2024stochastic}. A lighter-weight variant suited to settings where labels are unavailable is test-time adaptation \cite{wang2021tent}, which updates normalization statistics directly from unlabeled test batches via entropy minimization. In resource-constrained edge deployments, selective layer adaptation and lightweight fine-tuning further reduce the computational cost of continual learning \cite{daniels2023efficient}.

Despite these advances, existing continual learning methods share two fundamental assumptions. First, they assume that online adaptation can temporarily interrupt the primary inference task without violating application-level timing requirements. Such an assumption is unsuitable for mission-critical edge AI systems that require continuous real-time inference while adaptation proceeds in the background. Second, they assume reliable hardware execution throughout the adaptation process. When a transient hardware fault corrupts a parameter or gradient update, the resulting error becomes indistinguishable from a valid optimization step and permanently contaminates the learned model. Consequently, no existing regularization method, replay strategy, parameter isolation technique, or test-time adaptation framework provides a mechanism to simultaneously preserve uninterrupted inference, support continual adaptation, and contain hardware-induced parameter corruption.

\section{Related Work}\label{sec:relatedwork}

\subsection{Continual Learning}

Continual learning enables deployed models to adapt to evolving data distributions while mitigating catastrophic forgetting \cite{kirkpatrick2017overcoming}. Existing approaches include regularization-based methods such as Elastic Weight Consolidation (EWC) \cite{kirkpatrick2017overcoming} and Synaptic Intelligence \cite{zenke2017synaptic}, rehearsal-based techniques such as iCaRL~\cite{rebuffi2017icarl}, and knowledge-distillation methods such as Learning without Forgetting (LwF)~\cite{li2018lwf}, parameter-isolation methods \cite{christophides2024stochastic}, and lightweight adaptation strategies such as Test-Time Adaptation \cite{wang2021tent} and selective layer fine-tuning for edge devices \cite{daniels2023efficient}. Mateo-Garcia \textit{et al.} \cite{mateo2023orbit} further demonstrate the feasibility of onboard continual learning for satellite applications. Although these methods effectively address distribution shift, they assume reliable hardware execution and typically perform adaptation directly on the inference model. Consequently, they neither preserve uninterrupted real-time inference during adaptation nor provide mechanisms for containing hardware-induced corruption throughout the learning process.

\subsection{Fault-Tolerant Neural Networks}

Hardware reliability of deep neural networks has been extensively investigated through statistical fault injection, vulnerability characterization, and selective protection techniques. Statistical fault injection frameworks \cite{leveugle2009statistical,ruospo2025quantitative} enable scalable reliability evaluation, while accelerator-level studies demonstrate that fault propagation strongly depends on the underlying hardware architecture \cite{reagen2018ares}. For Vision Transformers, prior work identifies highly vulnerable architectural components and proposes selective hardening strategies based on vulnerability profiling \cite{xue2023softerror,he2025fine,liao2025analyzing,friasdominguez2026dependability}. Hardware-oriented protection mechanisms further reduce redundancy overhead through selective modular redundancy and lightweight fault-tolerance schemes \cite{tedeschi2025safeneureka,shao2024spaceborne,nazari2024fortune}. However, these techniques assume static model parameters and fixed operating conditions. Protection policies are derived offline and remain unchanged throughout deployment, making them unsuitable for continuously adapting models operating under varying reliability requirements.


\section{Methodology}\label{sec:methodology}

\begin{figure*}[t]
    \centering
    \includegraphics[width=\textwidth]{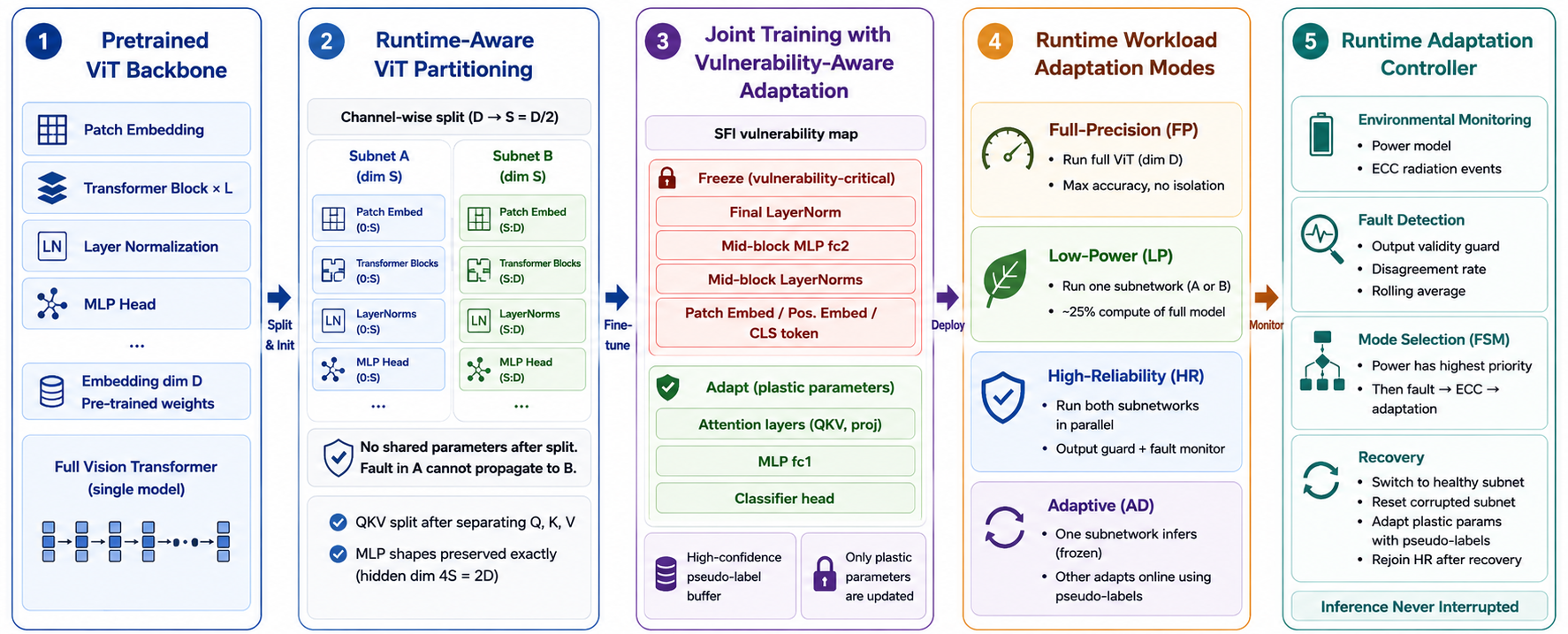}
    \caption{The five-stage WARD pipeline.}
    \label{fig:methodology}
\end{figure*}

\subsection{Overview of WARD}

WARD is a runtime-adaptive inference framework that continuously adjusts the execution strategy of a Vision Transformer according to changing operating conditions while maintaining uninterrupted inference. Rather than relying on a fixed execution pipeline throughout deployment, the proposed framework dynamically balances inference accuracy, computational cost, reliability, and online adaptation based on the current workload and environmental conditions. Figure~\ref{fig:methodology} illustrates the complete WARD workflow.

The framework is driven by three runtime signals. The first is the available computational budget, determined by the system power state. The second is the hardware reliability condition, obtained from onboard error monitoring and runtime fault detection. The third is the observed input distribution, which determines whether continual adaptation is required. These signals are continuously evaluated by a runtime controller that selects one of four operating modes to satisfy the current system requirements.

The four operating modes provide complementary execution strategies. \textit{Full-Precision} mode executes the original Vision Transformer to maximize inference accuracy under benign operating conditions. \textit{Low-Power} mode executes a single lightweight subnetwork to minimize computational cost when available resources become limited. \textit{High-Reliability} mode simultaneously executes two independent subnetworks to tolerate hardware faults while maintaining uninterrupted inference. Finally, \textit{Adaptive} mode performs continual learning on one subnetwork while the second continues serving inference, allowing the model to adapt to distribution shifts without interrupting real-time operation.

To demonstrate the practical deployability of the proposed framework, WARD is implemented on a programmable AI accelerator extended with lightweight runtime support for mode scheduling and resource management. The hardware implementation does not introduce a new accelerator architecture but instead demonstrates that the proposed runtime framework can be integrated into existing programmable AI accelerators through minimal architectural modifications.

\subsection{Runtime-Aware ViT Partitioning}

The runtime adaptation mechanism of WARD is enabled through a channel-wise partitioning of a pretrained Vision Transformer into two physically independent subnetworks. Rather than sharing parameters throughout execution, every parameter tensor is divided into two non-overlapping channel groups assigned to subnet~A and subnet~B. Consequently, each subnetwork maintains an independent parameter storage and execution path, allowing the runtime controller to selectively activate, adapt, or replicate computation according to the selected operating mode while preventing hardware faults from propagating between subnetworks.

Let $D$ denote the embedding dimension of the original Vision Transformer and $S=D/2$ the embedding dimension of each subnetwork. Every parameter tensor is partitioned along its channel dimension such that the two subnetworks collectively preserve the complete representational capacity of the original model without sharing parameters. After partitioning, both subnetworks are instantiated as independent \texttt{VisionTransformer} modules and jointly fine-tuned on the target task to compensate for the reduced embedding dimension while maintaining complementary feature representations.

The only layer requiring a dedicated partitioning strategy is the combined query-key-value (QKV) projection. Standard Vision Transformer implementations store the three projections as a single matrix
$\mathbf{W}_{QKV}\in\mathbb{R}^{3D\times D}$.
Directly partitioning this matrix would mix attention heads across subnetworks and violate attention consistency. Therefore, the Q, K, and V projections are first separated before channel-wise partitioning.

Using $[\mathbf{M}]_{p:q,r:s}$ to denote the submatrix spanning rows $p$ to $q-1$ and columns $r$ to $s-1$, the query projection for each subnetwork is obtained as

\begin{equation}
\mathbf{W}^{(A)}_Q=[\mathbf{W}_Q]_{0:S,0:S},
\qquad
\mathbf{W}^{(B)}_Q=[\mathbf{W}_Q]_{S:D,S:D},
\label{eq:qsplit}
\end{equation}

with identical partitioning applied to $\mathbf{W}_K$ and $\mathbf{W}_V$. The resulting subnet-specific QKV matrices become

\begin{equation}
\mathbf{W}^{(A)}_{QKV}
=
\left[
\mathbf{W}^{(A)}_Q
\;\Vert\;
\mathbf{W}^{(A)}_K
\;\Vert\;
\mathbf{W}^{(A)}_V
\right]
\in\mathbb{R}^{3S\times S},
\label{eq:qkv}
\end{equation}

and equivalently for subnet~B.

The multilayer perceptron (MLP) layers naturally preserve their structure after channel partitioning. Since the original hidden dimension expands from $D$ to $4D$, each subnetwork maintains a hidden dimension of $4S$, producing exact tensor dimensions without interpolation or approximation. The same channel-wise partitioning is applied to the patch embedding, positional embedding, class token, Layer Normalization parameters, and classifier head.

Following initialization from the partitioned weight dictionaries, both subnetworks are jointly fine-tuned on the target dataset. Independent forward passes are executed through subnet~A and subnet~B, their cross-entropy losses are accumulated, and gradients are independently propagated through each subnetwork. Joint optimization allows both subnetworks to specialize their respective feature representations while preserving the overall discriminative capability of the original Vision Transformer. More importantly, the resulting physical separation establishes the foundation for the runtime operating modes presented in the following subsection, enabling independent inference, redundancy, and continual adaptation without interrupting the primary inference task.

\subsection{Runtime Operating Modes}

The proposed channel-wise partitioning enables WARD to dynamically adapt both the computational workload and the reliability strategy according to changing runtime conditions. Rather than executing a fixed inference pipeline throughout deployment, the runtime controller continuously selects one of four operating modes based on the available power budget, hardware reliability condition, and adaptation requirements. Each mode represents a different trade-off between inference accuracy, computational cost, reliability, and online learning while maintaining uninterrupted execution.

\subsubsection{Full-Precision Mode}

Full-Precision (FP) mode executes the original pretrained Vision Transformer without partitioning, providing the highest inference accuracy during benign operating conditions where neither energy constraints nor elevated hardware fault rates are present. Since the complete network executes as a single model, no redundancy or fault isolation is available, and a hardware fault affecting any model parameter directly propagates to the final prediction. Consequently, FP mode serves as the baseline operating mode whenever maximum prediction accuracy is prioritized over computational efficiency or fault tolerance.

\subsubsection{Low-Power Mode}

Low-Power (LP) mode minimizes computational cost by executing only a single WARD subnetwork. Reducing the embedding dimension from $D$ to $S=D/2$ decreases the computational complexity of the self-attention operation to

\[
\left(\frac{S}{D}\right)^2 = 0.25
\]

of the original model, while the computational cost of the MLP layers decreases proportionally. This mode targets power-constrained operating conditions where maintaining uninterrupted inference becomes more important than maximizing accuracy or fault tolerance. Since only one subnetwork is active, no redundant execution path exists and hardware faults directly affect the inference result.

\subsubsection{High-Reliability Mode}

High-Reliability (HR) mode increases fault tolerance by executing both subnetworks concurrently. Under fault-free operation, the prediction logits produced by subnet A and subnet B are averaged,

\begin{equation}
\hat{\mathbf{l}}
=
\frac{\mathbf{l}_A+\mathbf{l}_B}{2},
\label{eq:avg}
\end{equation}

thereby exploiting the complementary feature representations learned during joint training.

To guarantee uninterrupted inference, each output is simultaneously verified for numerical validity. Let $v_X=1$ denote that subnet $X$ produces only finite output values and $v_X=0$ otherwise. The final prediction is determined according to

\begin{equation}
\hat{\mathbf{l}}=
\begin{cases}
(\mathbf{l}_A+\mathbf{l}_B)/2, & v_A=1,\;v_B=1,\\[3pt]
\mathbf{l}_B, & v_A=0,\;v_B=1,\\[3pt]
\mathbf{l}_A, & v_A=1,\;v_B=0,\\[3pt]
\mathbf{l}_A, & v_A=0,\;v_B=0.
\end{cases}
\label{eq:guard}
\end{equation}

Consequently, a hardware fault affecting one subnetwork does not interrupt the primary inference task as long as the second execution path remains operational. The physical isolation established during the partitioning stage therefore provides runtime fault containment without requiring complete model replication.

\subsubsection{Adaptive Mode}

Adaptive (AD) mode addresses gradual input distribution shifts while preserving uninterrupted real-time inference. One subnetwork remains frozen and continuously performs inference, whereas the second subnetwork is updated online using high-confidence pseudo-labels generated by the inference subnetwork. A sample $(\mathbf{x}_i,\hat{y}_i)$ is admitted to the adaptation buffer $\mathcal{B}$ when

\begin{equation}
(\mathbf{x}_i,\hat{y}_i)\in\mathcal{B}
\iff
\max_c p_{A,i,c}\ge\tau,
\qquad
\hat{y}_i=\arg\max_c p_{A,i,c},
\label{eq:pseudo}
\end{equation}

where $\tau$ denotes the confidence threshold.

Online adaptation is restricted to the fault-resilient parameter subset

\begin{equation}
\Theta_{\mathrm{plastic}}
=
\Theta
\setminus
\Theta_{\mathrm{vuln}},
\label{eq:plastic}
\end{equation}

where $\Theta_{\mathrm{vuln}}$ represents the vulnerability-critical parameters identified through offline statistical fault injection analysis. Consequently, reliability-critical parameters remain unchanged throughout adaptation while the remaining parameters continuously compensate for distribution shift. Because inference and adaptation execute on physically independent subnetworks, WARD maintains uninterrupted real-time operation throughout the learning process, eliminating the need to suspend inference while updating the deployed model.

\subsection{Runtime Adaptation Controller}

The Runtime Adaptation Controller continuously monitors the execution environment and dynamically selects the most appropriate operating mode according to the current system conditions. Rather than executing a fixed inference pipeline throughout deployment, WARD continuously balances computational efficiency, reliability, and online adaptation using three runtime signals: available power, hardware reliability, and model behaviour. Figure~\ref{fig:methodology} illustrates the interaction between the controller, the operating modes, and the underlying accelerator.

\subsubsection{Environmental Monitoring}

Runtime adaptation begins by continuously monitoring the available computational resources and the hardware operating condition. Available power is estimated using a continuous orbital energy model that combines instantaneous solar generation with battery state-of-charge.

Let $b_t$ denote the battery charge at time step $t$, $C_b$ the battery capacity, $s_t$ the available solar energy, $\alpha$ the charging efficiency, $\beta$ the baseline platform consumption, and $u_t$ the computational cost of the currently active operating mode. The battery evolution is

\begin{equation}
b_{t+1}
=
\min
\left(
b_t+\alpha s_t-\beta-u_t,\;
C_b
\right),
\label{eq:battery}
\end{equation}

where computationally intensive operating modes consume proportionally more energy. A power-availability signal is asserted only when both the available solar input and battery charge exceed predefined operating thresholds.

In parallel, the controller continuously receives hardware reliability information from the onboard ECC monitor, which reports elevated radiation activity whenever correctable memory errors are detected. Together, the power and reliability monitors provide the environmental information required for runtime workload adaptation.

\subsubsection{Fault Detection}

While ECC events indicate elevated radiation exposure, they do not necessarily imply that inference has been corrupted. WARD therefore performs runtime fault detection directly from the network outputs using two complementary mechanisms.

The first verifies the numerical validity of each subnetwork output by detecting non-finite values generated during inference. The second evaluates the prediction disagreement between the two physically independent subnetworks. The per-batch disagreement rate is

\begin{equation}
r_t
=
\frac{1}{|\mathcal{X}|}
\sum_{i\in\mathcal{X}}
\mathbf{1}
\left[
\arg\max
\mathbf{l}_i^A
\neq
\arg\max
\mathbf{l}_i^B
\right],
\label{eq:disagree_rate}
\end{equation}

where $|\mathcal{X}|$ denotes the batch size.

To suppress transient prediction variations, disagreement measurements are accumulated within a rolling observation window $\mathcal{W}$. A hardware fault is declared whenever the rolling disagreement exceeds

\begin{equation}
\bar r
=
\frac{1}{|\mathcal{W}|}
\sum_{r_i\in\mathcal{W}}
r_i
>
\theta_d,
\label{eq:roll}
\end{equation}

where $\theta_d$ is determined from the healthy baseline disagreement observed during fault-free execution. Whenever a non-finite output is detected, the disagreement is immediately set to one, allowing catastrophic failures to be identified without waiting for the rolling window to converge. A parallel single-step check evaluates $r_t > \theta_d$ directly at each inference step, independent of the rolling window accumulation. When a hardware fault produces non-finite outputs, $r_t$ is set to 1.000 and this direct check flags the fault on the same step without waiting for the window to fill. The rolling window provides detection for faults whose disagreement accumulates gradually; the direct check handles catastrophic failures that cause complete prediction divergence on the first affected batch. The 7-step detection latency reported in Table~\ref{tab:fault} reflects the rolling window mechanism under the isolated fault injection experiment. The fault in the orbital simulation (Section~\ref{sec:experiments}) is detected at step~0 through the direct per-step check, since $r_t = 1.000$ immediately exceeds $\theta_d$.

\subsubsection{Mode Selection}

The runtime controller evaluates the monitored environmental conditions together with the fault status to determine the operating mode executed during the next inference cycle.

Let $\text{pow}$ denote the power-availability signal, $\text{ecc}$ the hardware reliability monitor, $\text{fault}$ the runtime fault detector, and $c$ the number of consecutive fault-free execution steps. Mode selection follows

\begin{equation}
\text{mode}_t
=
\begin{cases}
\text{low-power},
&
\text{pow}=0,
\\[2pt]


\text{high-reliability},
&
\text{pow}=1,\;
\text{fault}=0,\;
\text{ecc}=1,
\\[2pt]

\text{adaptive},
&
\text{pow}=1,\;
\text{fault}=0,\;
\text{ecc}=0,\;
c\ge c_{\min},
\\[2pt]

\text{full-precision},
&
\text{otherwise.}
\end{cases}
\label{eq:fsm}
\end{equation}

Power availability receives the highest priority because computational constraints override all other objectives. Confirmed hardware faults immediately activate recovery, while elevated radiation conditions trigger High-Reliability mode to increase fault tolerance. Adaptive mode is entered only after the system remains fault-free for a sufficiently long observation period, ensuring that online learning begins only under stable operating conditions.

\subsubsection{Recovery}

Whenever a hardware fault is confirmed, WARD immediately redirects inference to the healthy subnetwork while restoring the affected execution path. The corrupted subnetwork is first reset to its clean checkpoint before deployment. A confidence-filtered pseudo-label buffer is then generated using the healthy subnetwork according to Eq.~(\ref{eq:pseudo}). If sufficient high-confidence samples are available, only the plastic parameter subset defined in Eq.~(\ref{eq:plastic}) is updated while the reliability-critical parameters remain frozen. After adaptation completes, the recovered subnetwork rejoins execution in High-Reliability mode, allowing the disagreement monitor to verify correct operation before normal runtime adaptation resumes. Throughout the entire recovery procedure, inference continues uninterrupted using the healthy execution path.

\subsection{Hardware Realization}

To demonstrate the practical deployability of the proposed runtime framework, WARD is implemented on a programmable FPGA-based AI accelerator extended in this work, with lightweight runtime control support called PERUN. The hardware implementation maps the four runtime operating modes directly onto the accelerator runtime controller. 

As illustrated in Fig. 4, PERUN integrates a Rocket Chip RISC-V processor that serves as the runtime management unit. The processor communicates with external devices to receive commands from the system master and to acquire environmental information used for runtime mode selection. It is connected to external DDR memory through a memory interface controller, where all network parameters, runtime configurations, and accelerator programs are stored. Communication among the Rocket processor, the PERUN accelerator, and the external memory is realized through an AXI interconnect, which provides a unified communication fabric for processor-memory, processor-accelerator, and accelerator-memory transactions.
The runtime controller supports four interrupt sources corresponding to the predefined operating modes: Full-Performance (FP), Low-Power (LP), High-Resilience (HR), and Adaptive (AD). Upon receiving an interrupt, the Rocket processor invokes the corresponding interrupt service routine (ISR), which reconfigures PERUN by updating its control registers through the AXI interface. The ISR subsequently loads the required instruction stream into the Code RAM (CRAM) and the associated connectivity information into the Link RAM (LRAM), according to the selected operating mode, before issuing the execution command to the accelerator. This software-assisted runtime management enables rapid transitions among operating modes while requiring only lightweight processor intervention.

In Full-Precision mode, the complete Vision Transformer executes using the standard execution flow. Low-Power mode activates only a single subnetwork to reduce computational workload and energy consumption. High-Reliability mode executes both subnetworks concurrently while enabling runtime output comparison and fault monitoring. Adaptive mode maintains inference on one execution path while the second path performs continual learning in the background, allowing uninterrupted inference throughout online adaptation.

\begin{figure}[t]
    \centering
    \includegraphics[width=\columnwidth]{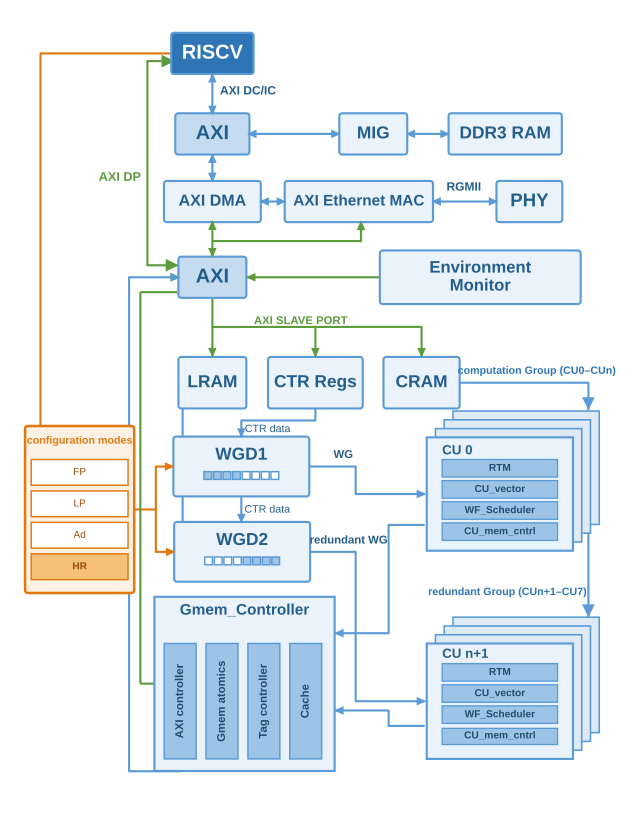}
    \caption{Hardware architecture of the PERUN accelerator in HR configuration.}
    \label{fig:hardware_architecture}
\end{figure}

\begin{figure}[t]
    \centering
    \includegraphics[width=\columnwidth]{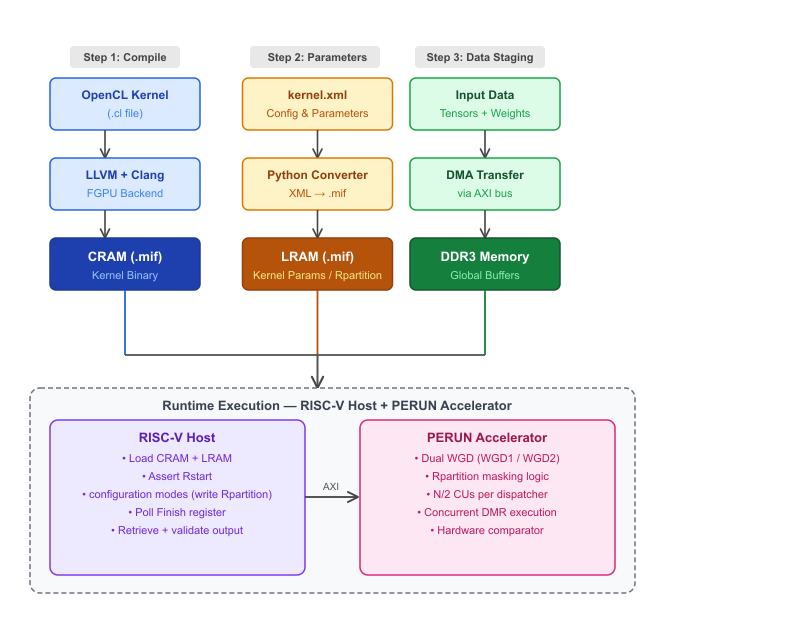}
    \caption{Compilation and runtime deployment flow for PERUN.}
    \label{fig:hardware_flow}
\end{figure}
Supporting these execution modes requires only lightweight runtime extensions to the baseline architecture. The runtime controller manages operating-mode selection, execution scheduling, memory allocation, and synchronization between the two subnetworks without modifying the underlying processing elements or computational datapath. Since the proposed framework operates entirely at the runtime scheduling layer, it remains largely independent of the underlying accelerator organization and can therefore be integrated into a broad class of programmable AI accelerators with minimal architectural modifications.

The complete deployment flow is illustrated in Fig.~\ref{fig:hardware_flow}. Offline, the Vision Transformer model, runtime parameters, and WARD operating modes are compiled and transferred to the accelerator memories. During execution, the runtime controller continuously monitors the environmental signals and dynamically schedules the appropriate operating mode while coordinating inference, continual learning, and fault recovery. This organization demonstrates that WARD extends existing programmable AI accelerators through lightweight runtime support rather than extensive hardware redesign.

\section{Experimental Results}
\label{sec:experiments}

\subsection{Experimental Setup}

The proposed WARD framework is evaluated from three complementary perspectives: (i) the effectiveness of the four runtime operating modes under varying deployment conditions, (ii) the ability of the runtime adaptation framework to maintain dependable inference during hardware faults and distribution shifts, and (iii) the practicality of deploying the proposed framework on a programmable AI accelerator with lightweight architectural support.

All software experiments use ViT-Tiny~\cite{dosovitskiy2021vit} pretrained on ImageNet-21K~\cite{ridnik2021} and fine-tuned on EuroSAT~\cite{helber2019eurosat}, a 10-class satellite image classification benchmark containing approximately 27,000 images. The proposed channel-wise partitioning produces two independent 96-dimensional subnetworks that are jointly fine-tuned for three epochs using Adam with an initial learning rate of $10^{-4}$ and a cosine learning-rate schedule. Distribution shift is emulated using a mild ColorJitter augmentation (brightness 0.1, contrast 0.1, saturation 0.1, and hue 0.05), representing gradual sensor variation, illumination changes, and seasonal appearance changes typically encountered during long-term edge deployment.

Reliability evaluation follows the statistical fault injection methodology introduced by Leveugle \textit{et al.}~\cite{leveugle2009statistical}. Unless otherwise stated, faults are injected into the deployed subnet during inference, while recovery is performed using the runtime mechanisms presented in Section~\ref{sec:methodology}. Latency is measured by averaging twenty forward passes on a CPU platform, reflecting the computational constraints of resource-limited edge processors. Four runtime operating modes are evaluated throughout this section: Full-Precision (FP), Low-Power (LP), High-Reliability (HR), and Adaptive (AD).

To demonstrate the practical deployability of the proposed framework, WARD is further implemented on the programmable PERUN accelerator described in Section~\ref{sec:methodology}. Rather than evaluating the accelerator itself, the hardware experiments quantify the additional runtime overhead introduced by WARD and demonstrate that the proposed runtime framework can be integrated through lightweight scheduling-level extensions without modifying the computational datapath.

\subsection{Runtime Operating Modes}

\begin{figure*}[t]
    \centering
    \includegraphics[width=\textwidth]{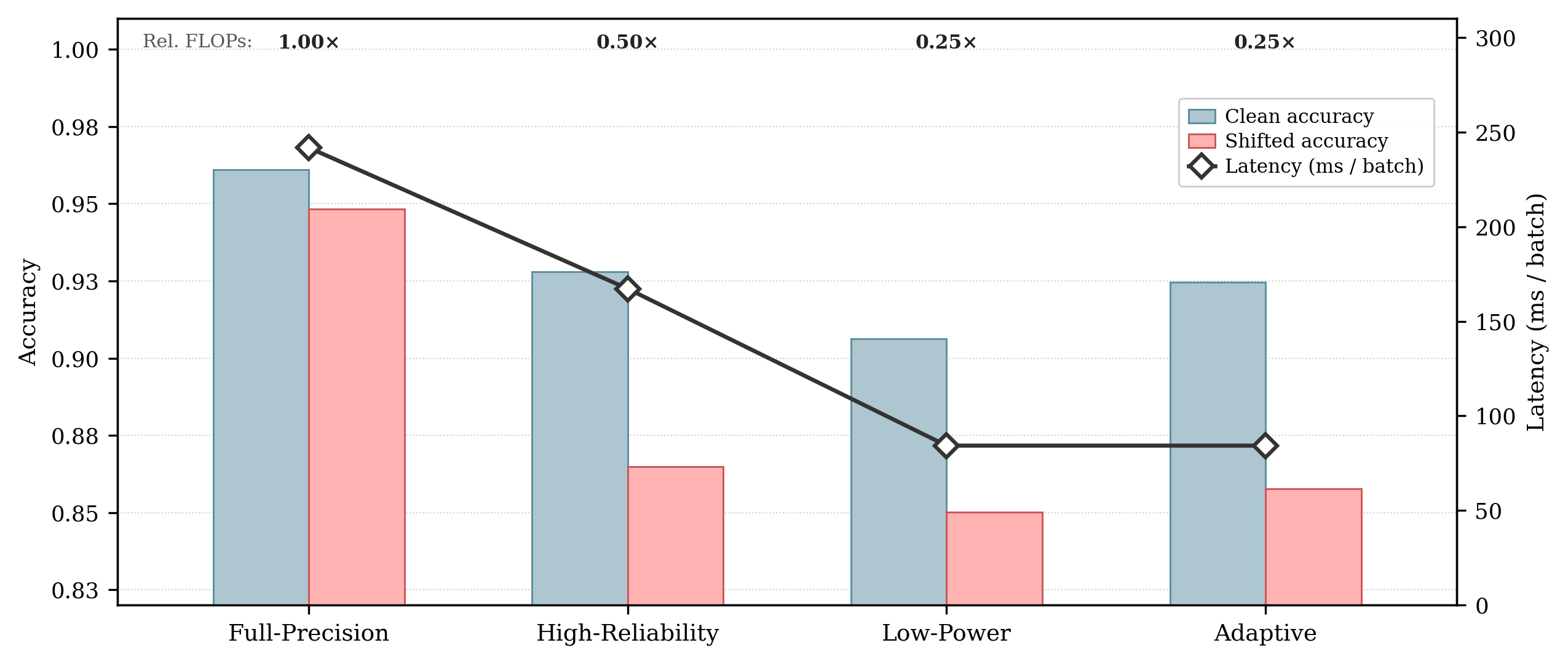}
    \caption{Accuracy and computational overhead of the four WARD operating modes.}
    \label{fig:modes}
\end{figure*}

The proposed runtime framework dynamically selects one of four operating modes according to the available power budget, hardware reliability condition, and adaptation requirements. Table~\ref{tab:modes} summarizes the overall performance of the four operating modes, while Fig.~\ref{fig:modes} illustrates the corresponding trade-offs between inference accuracy and computational cost.

\begin{table*}[t]
\caption{Performance comparison of the proposed runtime operating modes.}
\label{tab:modes}
\centering
\begin{tabular}{lccccc}
\hline
Mode & Clean Accuracy & Shifted Accuracy & Latency (ms/batch) & Relative Parameters & Relative FLOPs \\
\hline
Full Precision (FP)      & 0.9611 & 0.9483 & 241.7 & 1.00 & 1.00 \\
High Reliability (HR)    & 0.9280 & 0.8648 & 167.2 & 0.52 & 0.50 \\
Low Power (LP)           & 0.9063 & 0.8502 & 84.3  & 0.26 & 0.25 \\
Adaptive (AD)            & 0.9246 & 0.8576 & 84.3  & 0.52 & $0.25^{\dagger}$ \\
\hline
\multicolumn{6}{l}{$^{\dagger}$Serving FLOPs only. The second subnetwork performs continual learning in the background.}\\
\multicolumn{6}{l}{Latency values are averaged over twenty forward passes on CPU.}
\end{tabular}
\end{table*}

\subsubsection{Full-Precision Mode}

Full-Precision (FP) mode represents the highest-performance operating point of the proposed framework and serves as the reference configuration for all subsequent comparisons. The complete 192-dimensional Vision Transformer executes without partitioning, achieving a clean Top-1 accuracy of 96.11\% and a shifted accuracy of 94.83\%, as reported in Table~\ref{tab:modes}. This performance comes at the highest computational cost, requiring the complete parameter budget and the full computational workload of the original model, resulting in an average latency of 241.7\,ms per batch.

Although FP mode provides the highest inference accuracy, it offers no runtime fault tolerance. Any hardware fault affecting the deployed model immediately propagates to the inference output since no redundant execution path or recovery mechanism is available. Consequently, FP mode is intended for benign operating conditions where maximum predictive accuracy outweighs power efficiency and runtime reliability.

\subsubsection{Low-Power Mode}

Low-Power (LP) mode minimizes computational workload by executing only a single WARD subnetwork. Reducing the embedding dimension from 192 to 96 lowers the computational complexity of the attention and MLP layers to approximately one quarter of the original model, resulting in only 0.25 relative FLOPs and an average latency of 84.3\,ms per batch.

Despite reducing the active parameter budget to only 26\% of the original model, LP mode maintains a clean accuracy of 90.63\% and a shifted accuracy of 85.02\%, representing only a modest reduction compared with FP mode while substantially lowering computational demand. LP mode therefore provides the preferred operating point during power-constrained mission phases where maintaining uninterrupted inference is more important than maximizing prediction accuracy.

Since only one execution path remains active, LP mode does not provide runtime redundancy. Hardware faults therefore directly affect the inference output and require a transition to the High-Reliability mode before recovery can begin.

\subsubsection{High-Reliability Mode}

High-Reliability (HR) mode activates both subnetworks simultaneously and combines their predictions through the output voting mechanism introduced in Section~\ref{sec:methodology}. Under fault-free execution, the two subnetworks produce a voted clean accuracy of 92.80\% and a shifted accuracy of 86.48\%, while requiring approximately half of the original computational workload and an average latency of 167.2\,ms per batch.

The primary objective of HR mode is dependable inference rather than maximum accuracy. When a single weight fault is injected into subnet A, the corrupted subnet immediately degrades to 11.06\% accuracy, whereas the voted system output remains at 92.17\%, effectively preserving the inference performance of the healthy subnet. The disagreement monitor detects the fault after seven execution steps, exceeding the predefined disagreement threshold and initiating the runtime recovery procedure. Following checkpoint restoration and continual adaptation, the recovered system achieves a voted accuracy of 92.67\%.

\begin{figure*}[t]
    \centering
    \includegraphics[width=\textwidth]{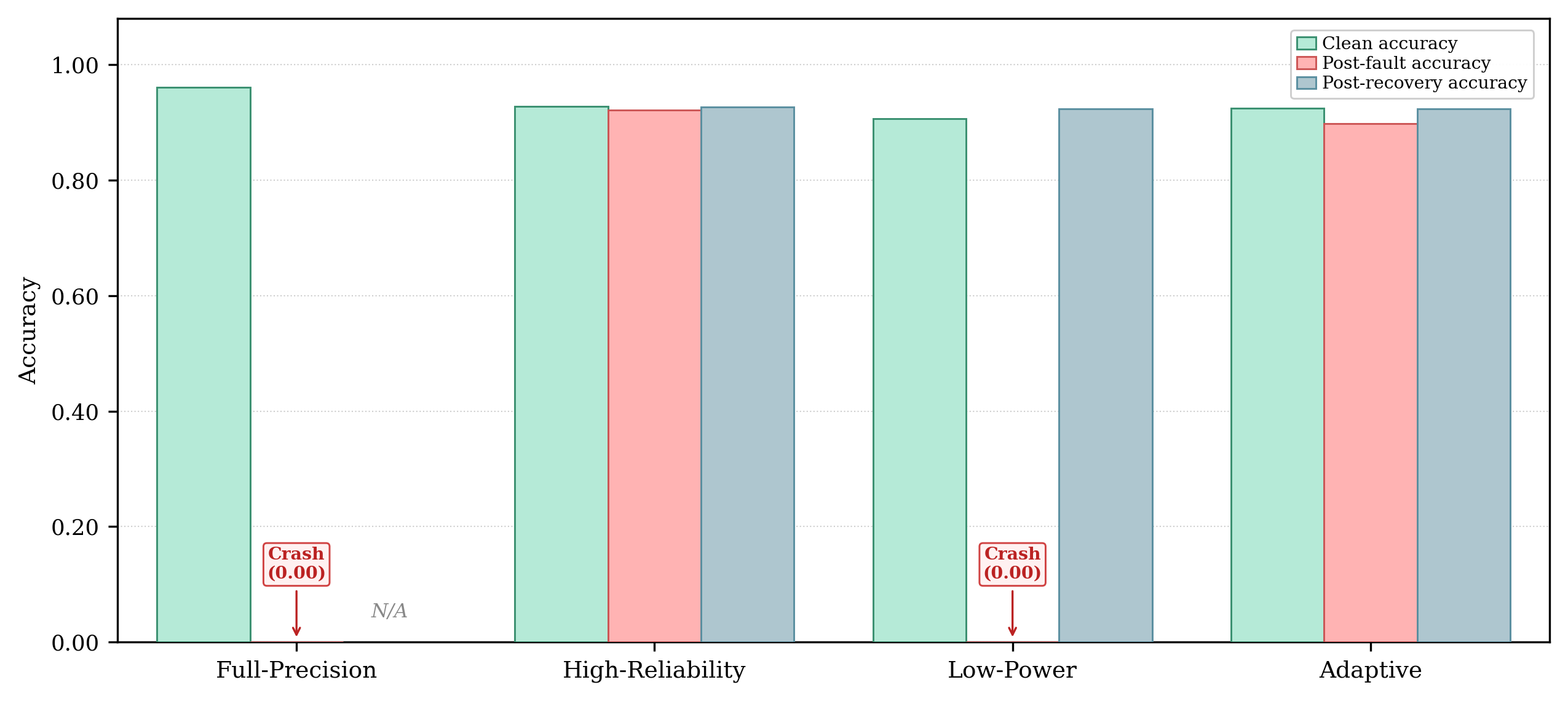}
    \caption{Fault response of the four runtime operating modes.}
    \label{fig:fault}
\end{figure*}

Figure~\ref{fig:fault} compares the fault response of all four operating modes, while Table~\ref{tab:fault} summarizes the corresponding recovery characteristics.

\begin{table}[t]
\caption{Fault handling capability of the proposed runtime operating modes.}
\label{tab:fault}
\centering
\begin{tabular}{lcccc}
\hline
Mode & Post-fault & Detection & Steps$^{*}$ & Post-recovery \\
\hline
FP & 0.0000 & None         & N/A & N/A    \\
HR & 0.9217 & Disagreement &   7 & 0.9267 \\
LP & 0.0000 & Output fault &   0 & 0.9235 \\
AD & 0.8981 & Fault + fallback & 0 & 0.9239 \\
\hline
\multicolumn{5}{l}{\parbox{0.85\columnwidth}{\footnotesize
$^{*}$Steps via rolling disagreement window. 0 = detected on
the first affected batch through the direct per-step check
($r_t > \theta_d$).}}
\end{tabular}
\end{table}

\subsubsection{Adaptive Mode}

Adaptive (AD) mode enables continual learning without interrupting the primary inference task. During execution, subnet A continuously serves inference while subnet B adapts to the incoming data distribution using high-confidence pseudo-labels with a confidence threshold of 0.88. Since only subnet A lies on the critical inference path, the serving latency remains identical to LP mode at 84.3\,ms per batch while maintaining only 0.25 relative serving FLOPs.

After online adaptation, AD mode achieves a clean accuracy of 92.46\% and a shifted accuracy of 85.76\%, improving upon the LP baseline while preserving identical inference latency. This demonstrates that continual learning progressively compensates for distribution shift without increasing the runtime cost of inference.

When a hardware fault occurs during adaptation, the runtime controller immediately redirects inference to the healthy execution path, maintaining an output accuracy of 89.81\% while the affected subnet is restored and re-adapted. Following recovery, the system returns to 92.39\% accuracy. To avoid adapting under unstable operating conditions, AD mode is activated only after twenty-five consecutive fault-free execution steps, ensuring that continual learning proceeds exclusively during reliable operating periods.

\subsection{Runtime Reliability and Adaptation}

The previous section demonstrates how the four operating modes balance computational cost, reliability, and adaptation under different deployment conditions. This section evaluates the reliability characteristics of the proposed WARD architecture itself. Specifically, three aspects are investigated: (i) whether the channel-wise partition preserves the fault behavior of the original Vision Transformer, (ii) whether the proposed vulnerability-aware parameter protection successfully balances reliability and adaptability, and (iii) whether the offline vulnerability characterization effectively reduces the fault-monitoring space during deployment.

\subsubsection{Statistical Fault Injection Validation}

The proposed framework derives its protected parameter set from the statistical fault characterization presented in the companion study. Since WARD partitions the original network into two independent subnetworks, it is first necessary to verify that this partitioning does not fundamentally alter the underlying fault behavior.

Statistical fault injection is therefore performed on subnet A using the finite-population sampling methodology of Leveugle \textit{et al.}~\cite{leveugle2009statistical}. Using the previously established informed prior of approximately 3\% network-level failure probability reduces the required sample size from 2,401 injections (conservative $p=0.5$) to only 280 injections at 95\% confidence and 2\% margin of error, representing an 8--9$\times$ reduction in experimental effort.

The resulting network-level failure rate of WARD is 1.79\% with a 95\% confidence interval of [0.23\%, 3.34\%]. As shown in Table~\ref{tab:sfi}, this interval overlaps with the 99\% confidence interval reported for the original full Vision Transformer, confirming that the proposed channel-wise partitioning preserves the intrinsic fault behavior of the network despite reducing the embedding dimension by half.

\begin{table}[t]
\caption{Statistical fault injection comparison between WARD and the original Vision Transformer.}
\label{tab:sfi}
\centering
\begin{tabular}{lcc}
\hline
Metric & WARD Subnet A & Original ViT \\
\hline
Parameters          & 1,436,074 & 5,526,346 \\
Required injections & 280       & 16,538 \\
Failure rate        & 1.79\%    & 2.93\% \\
95\% CI             & [0.23\%, 3.34\%] & --- \\
99\% CI             & --- & [2.60\%, 3.27\%] \\
Confidence overlap  & \multicolumn{2}{c}{Yes} \\
Exponent MSB share  & 84.5\% & 87.4\% \\
Fault-space reduction & \multicolumn{2}{c}{307$\times$} \\
\hline
\end{tabular}
\end{table}

\begin{figure}[t]
    \centering
    \includegraphics[width=\columnwidth]{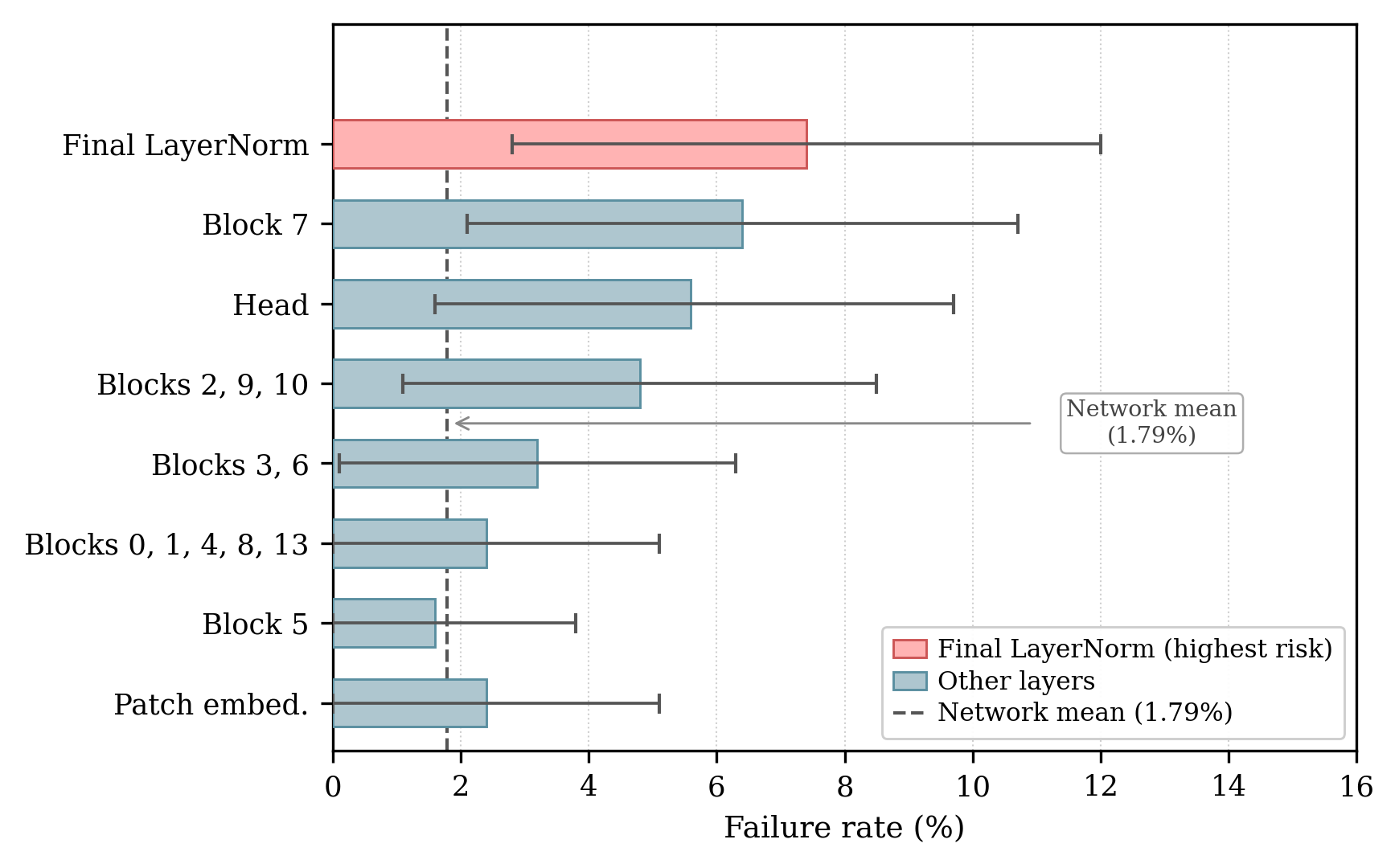}
    \caption{Layer-wise statistical fault injection failure rates for WARD subnet A. Error bars indicate the corresponding 95\% confidence intervals.}
    \label{fig:sfi}
\end{figure}

Figure~\ref{fig:sfi} further reports the layer-wise failure distribution obtained for WARD. Similar to the original model, the final Layer Normalization remains the most vulnerable component with a failure rate of 7.4\%, followed by encoder block~7 at 6.4\%. The patch embedding layer also exhibits an above-average failure rate of 2.4\%, reflecting its influence on all subsequent feature representations. The close agreement with the vulnerability distribution of the original network confirms that channel-wise partitioning preserves the critical reliability characteristics required by the proposed runtime framework. 

Furthermore, physical isolation between the two subnetworks is verified through byte-level comparison of every parameter tensor before and after fault injection. Across all evaluated scenarios, every parameter stored in subnet B remains identical to its pre-fault state following faults injected into subnet A. This confirms that the proposed channel-wise partitioning provides complete storage-level isolation between the two execution paths rather than merely logical separation through software abstraction.


\subsubsection{Reliability-Aware Parameter Protection}

The proposed continual learning framework updates only the parameter subset identified as fault-resilient during the offline vulnerability characterization. The remaining reliability-critical parameters remain frozen throughout online adaptation.

To validate this design choice, different protected parameter fractions ranging from 5\% to 50\% are evaluated. Across all evaluated configurations, the post-recovery accuracy varies only between 92.17\% and 92.98\%, corresponding to a total variation of only 0.81\%. This variation remains considerably smaller than the 3.31\% accuracy difference between the Full-Precision and High-Reliability operating modes, demonstrating that protecting a relatively small subset of reliability-critical parameters has only a minor influence on the achievable recovery performance.

The adaptation accuracy peaks near a protected fraction of 10\% (0.8446) and remains within
a 2.2\% range across the full evaluated sweep from 5\% to 50\%, with no monotone dependence
on the frozen fraction. The vulnerability-guided prefix list derived from the companion study targets approximately 16.9\% of the parameter space; because the list captures whole parameter groups rather than individual weights, the measured frozen fraction at runtime is 18.5\% (266,218 of 1,436,074 parameters per subnet). This operating point sits within 0.44\% of the adaptation accuracy peak without requiring explicit hyperparameter search. Consequently, the proposed parameter protection strategy preserves the reliability-critical regions of the model while maintaining sufficient adaptation capacity for continual learning.


\subsubsection{Fault-Space Reduction}

The vulnerability characterization further enables substantial reduction of the runtime fault-monitoring space. Rather than monitoring every possible parameter location, WARD focuses on the dominant failure mechanism identified during the offline analysis.

Restricting the monitored fault locations to the exponent most-significant bit within the twelve highest-vulnerability parameter groups reduces the number of candidate injection sites from 45,954,368 to only 149,568 locations. This corresponds to a 307$\times$ reduction in the explored fault space while still covering the dominant hardware failure mechanism.

This reduction substantially lowers the computational effort required for deployment-time reliability assessment and online health monitoring. Instead of evaluating the complete parameter space, only approximately 0.33\% of the original fault space must be considered while maintaining coverage of the overwhelming majority of observable failures. The proposed vulnerability-aware monitoring strategy therefore complements the runtime operating modes by enabling efficient reliability management without introducing excessive monitoring overhead.


\subsubsection{Random Bit-Flip Fault Model}
The fault injection experiments above use a worst-case targeted model in which faults are injected at the exponent most-significant bit of a randomly selected parameter. To assess whether the fault containment properties of the proposed modes generalize to the
full distribution of hardware upsets, 50 injections per mode were performed under a uniform random bit-flip model, in which the injected bit position is drawn uniformly from all 32 bit positions of a randomly selected parameter.

Under this model, HR and AD modes each record a 0\% failure rate across all 50 injections. The output voting guard and physical subnet isolation eliminate every user-visible output failure. FP mode produces a 2\% failure rate (1 of 50) and LP mode 4\% (2 of 50). These rates are consistent with the analytically predicted network-level failure probability under a uniform random flip, derived from the bit-field failure rate decomposition established in the companion study:

\begin{equation}
P(\text{fail} \mid \text{random})
\approx \tfrac{1}{32} \cdot 0.87 + \tfrac{7}{32} \cdot 0.01
      + \tfrac{23}{32} \cdot 0.001 \approx 3\%.
\label{eq:random_pred}
\end{equation}

This result confirms that the fault tolerance of HR and AD mode is not specific to the worst-case single-site injection but holds across the full distribution of uniformly random hardware upsets.

\subsection{Runtime Adaptation Controller}

\begin{figure*}[t]
    \centering
    \includegraphics[width=.8\textwidth]{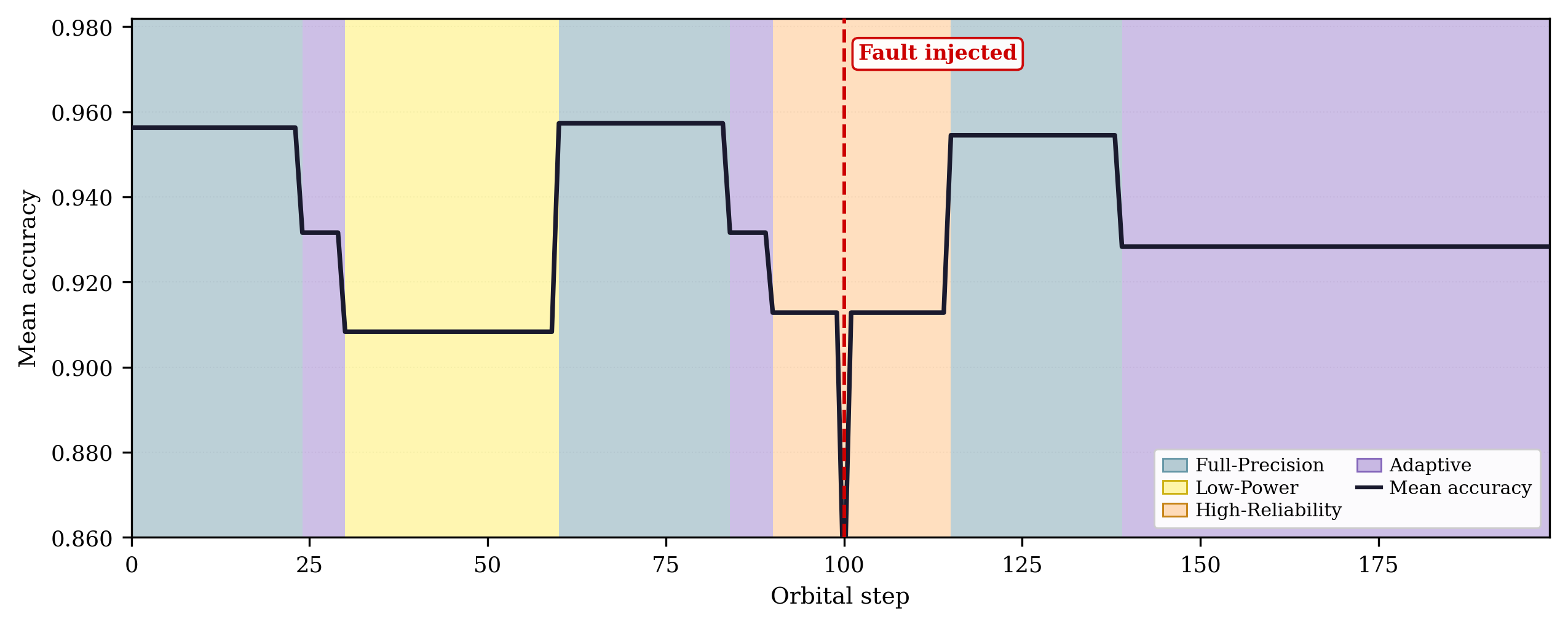}
    \caption{Operating-mode transitions during the 200-step orbital simulation.}
    \label{fig:orbital}
\end{figure*}

The runtime adaptation controller is evaluated over a 200-step orbital simulation incorporating the continuous solar power model, battery state-of-charge tracking, ECC-based radiation monitoring, and the finite-state machine described in Section~\ref{sec:methodology}. To assess the consequence of each mode transition, per-step Top-1 accuracy is evaluated against the labeled EuroSAT test set throughout the simulation; this is a post-hoc evaluation metric. In deployment the controller has no access to ground-truth labels and acts on the inter-subnet disagreement rate, the ECC monitor, and the available power state instead. Figure~\ref{fig:orbital} shows the resulting mode sequence.

The simulation opens under benign conditions, with solar generation above threshold and no ECC events active, so the controller selects Full-Precision mode. The mean Top-1 accuracy evaluated over the first 24 steps is 95.63\%. At step 24 the clean-step counter reaches the 25-step stability threshold and the controller enters Adaptive mode. Subnet A continues serving inference while subnet B begins background adaptation from confidence-filtered pseudo-labels; the mean evaluated accuracy over the transition window (steps 24--29) is 93.16\%.

Solar generation drops below the operating threshold at step 30 as the satellite enters eclipse. The controller switches to Low-Power mode, reducing the active workload to subnet A alone. Inference runs without interruption across the full eclipse period; mean evaluated accuracy over steps 30--59 is 90.83\%. Solar power resumes at step 60, Full-Precision mode re-engages, and after a further 25 consecutive fault-free steps Adaptive mode returns at step 84.

At step 90 the ECC monitor reports elevated radiation activity, consistent with a radiation belt entry, and the controller escalates to High-Reliability mode before any inference failure occurs. Both subnetworks begin running concurrently. A fault is injected at step 100 and the disagreement monitor registers a rate of 1.000, well above the 0.30 threshold, triggering recovery. Subnet B serves inference throughout while subnet A resets to its pre-fault checkpoint and re-adapts from the pseudo-label buffer. Recovery completes within a single simulation step and High-Reliability execution resumes. The mean evaluated accuracy across the combined interval (steps 90--114) is 91.00\%.

The ECC monitor clears at step 115 and the controller returns to Full-Precision mode; mean evaluated accuracy over the subsequent interval is 95.45\%. The clean-step counter reaches 25 at step 139 and Adaptive mode engages for the remainder of the simulation, ending at a mean evaluated accuracy of 92.83\%.

The simulation confirms that the controller handles compound orbital conditions correctly. Eclipse and radiation belt entries each trigger the appropriate mode transition, the fault injected at step 100 is contained and recovered within a single step without interrupting inference, and the post-recovery state retains High-Reliability rather than reverting to Full-Precision. The clean-step counter resets correctly after each environmental event, battery charge remains above the emergency threshold throughout, and all six state-machine validation conditions pass.

\subsection{Hardware Evaluation}

To demonstrate the practical applicability of the proposed runtime framework, WARD is implemented on the programmable PERUN accelerator using the lightweight architectural extensions described in Section~\ref{sec:methodology}. Unlike the previous software experiments, the objective of this evaluation is not to assess accelerator performance itself, but to quantify the implementation overhead required to support runtime operating modes, reliability-aware execution, and uninterrupted continual learning. Figure~\ref{fig:hardware_architecture} illustrates the modified accelerator organization, while Fig.~\ref{fig:hardware_flow} presents the complete compilation and deployment flow.

The proposed runtime framework requires only lightweight modifications to the baseline accelerator. As summarized in Table~\ref{tab:hw_overhead}, introducing the second runtime scheduler together with the associated control logic increases LUT utilization by approximately 2--3\%, while upgrading the runtime parameter memory to dual-port operation increases BRAM utilization by only 1--2\%. The remaining masking logic and hardware comparator contribute less than 1\% additional LUTs, resulting in an overall hardware overhead below 5\%.

Despite these architectural extensions, critical-path analysis confirms that the maximum operating frequency remains unchanged at approximately 240\,MHz for the accelerator domain. Consequently, WARD introduces runtime adaptability without affecting the timing characteristics or computational datapath of the baseline accelerator.

\begin{table}[t]
\caption{Hardware overhead introduced by WARD runtime support.}
\label{tab:hw_overhead}
\centering
\begin{tabular}{lc}
\hline
Metric & Value\\
\hline
Additional scheduler logic & 2--3\% LUT\\
Additional runtime memory & 1--2\% BRAM\\
Comparator and masking logic & $<1$\% LUT\\
Total hardware overhead & $<5$\%\\
Accelerator frequency & $\sim$240 MHz\\
\hline
\end{tabular}
\end{table}

\begin{table}[t]
\caption{Representative accelerator operating points for the WARD runtime modes.}
\label{tab:mode_accel_benchmarks}
\centering
\begin{tabular}{lccc}
\hline
Runtime Mode & Compute Units & Power (W) & Energy (J) \\
\hline
Full Precision (FP)   & 4 & 3.995 & 920.6 \\
Low Power (LP)        & 2 & 3.586 & 504.5 \\
High Reliability (HR) & 4 & 3.995 & 820.4 \\
\hline
\end{tabular}
\end{table}

Unlike Dynamic Partial Reconfiguration (DPR), which requires partial bitstream loading and execution interruption, WARD performs runtime adaptation entirely through software-programmable control registers. Switching between operating modes therefore requires only 50 clock cycles (small interrupt subroutine).

During High-Reliability mode, both subnetworks execute concurrently while their outputs are compared by lightweight hardware logic. Across all evaluated workloads, the measured execution overhead remains consistent, averaging approximately 1.96$\times$ (range 1.79$\times$--2.14$\times$). This closely matches the theoretical cost of duplicate execution, confirming that the comparator and synchronization logic introduce negligible additional runtime overhead beyond the duplicated computation itself.

Importantly, this overhead is incurred only while High-Reliability mode is active. During Full-Precision, Low-Power, and Adaptive modes, the accelerator executes without any redundancy penalty, allowing WARD to dynamically trade computational efficiency for reliability according to runtime requirements.

To characterize the hardware cost of the proposed runtime modes, representative accelerator operating points were selected for the Full-Precision (FP), Low-Power (LP), and High-Reliability (HR) execution modes. Rather than reporting network-specific measurements, the selected operating points represent the execution profiles of the runtime scheduler on the programmable accelerator.
\begin{table}[t]
\caption{Comparison with representative runtime reliability accelerators.}
\label{tab:comparison}
\centering
\footnotesize
\setlength{\tabcolsep}{3pt}
\begin{tabular}{lccc>{\centering\arraybackslash}p{2.2cm}}
\hline
Approach & Area Overhead & Runtime Switching & DMR Overhead \\
\hline
Static DMR \cite{7} & $>100$\% & Fixed & $\sim2\times$ \\
FlexGrip \cite{17} & 6--22\% & No & Software \\
HMR-NEureka \cite{21} & $\sim9$\% & Few cycles & $\sim2\times$ \\
Safe-NEureka \cite{22} & Low & $<400$ cycles & Dynamic \\
WARD (proposed) & $<5$\% & 50 cycles &
\makecell[c]{1.79--2.14$\times$\\(1.96$\times$ avg.)} \\
\hline
\end{tabular}
\end{table}

As shown in Table~\ref{tab:mode_accel_benchmarks}, FP mode utilizes four compute units to maximize throughput, resulting in a measured power consumption of 3.995~W and an energy consumption of 920.6~J. LP mode reduces the active hardware resources to two compute units, lowering both power and energy to 3.586~W and 504.5~J, respectively. HR mode activates redundant execution while maintaining four compute units, consuming the same power as FP mode but requiring 820.4~J for the evaluated workload. These representative operating points demonstrate that WARD dynamically maps runtime modes onto different hardware resource allocations, enabling the accelerator to trade performance, energy efficiency, and reliability according to runtime requirements.

Table~\ref{tab:comparison} compares the proposed implementation against representative runtime redundancy architectures.

Compared with static hardware DMR \cite{7}, which typically requires complete duplication of the compute fabric and more than 100\% additional area, WARD achieves runtime-selectable redundancy with less than 5\% additional hardware. Compared with recent dynamic redundancy architectures such as HMR-NEureka, which reports approximately 9\% hardware overhead, the proposed implementation reduces the additional hardware cost by almost half while providing comparable runtime mode-switch latency \cite{21, 22}.

Unlike software redundancy techniques, which typically incur 2.5--4$\times$ execution overhead because of host synchronization and software comparison, WARD performs output verification directly in hardware, maintaining an average runtime overhead of only 1.96$\times$. Furthermore, unlike domain-specific accelerators, the proposed implementation preserves complete OpenCL compatibility, allowing existing kernels to execute without modification while dynamically switching between runtime operating modes.

The hardware evaluation demonstrates that supporting the proposed runtime framework requires only lightweight scheduling-level extensions while preserving the original accelerator datapath and programming model. The measured hardware overhead remains below 5\%, runtime mode transitions complete within 50 cycles, and the average redundancy overhead closely matches the theoretical cost of duplicate execution. Together, these results demonstrate that WARD can be integrated into existing programmable AI accelerators with minimal architectural modifications, making the proposed runtime adaptation framework practical for real-world edge AI deployments.


\section{Conclusion}\label{sec:conclusion}

This paper presented WARD, a runtime workload-adaptive framework for dependable Vision Transformer deployment on programmable edge AI accelerators. By combining channel-wise network partitioning, runtime mode scheduling, and reliability-aware continual learning, WARD enables uninterrupted inference while dynamically balancing accuracy, computational cost, and reliability according to changing operating conditions. Experimental results demonstrate up to 96.11\% classification accuracy, 0.25$\times$ serving FLOPs in Low-Power mode, a network-level failure rate of only 1.79\%, a 307$\times$ reduction in the monitored fault space, and less than 5\% hardware overhead with runtime mode switching in only 50 clock cycles. These results demonstrate that WARD provides a practical and lightweight solution for runtime-adaptive, dependable edge AI deployment.

\section*{ACKNOWLEDGMENT}
\small
This work was supported in part by the Estonian Research Council grant PUT PRG1467 ``CRASHLESS'', EU Grant Project 101160182 ``TAICHIP'', and by the Federal Ministry of Research, Technology and Space of Germany (BMFTR) for supporting Edge-Cloud AI for DIstributed Sensing and COmputing (AI-DISCO) project (Project-ID ``16ME1127'').

\bibliographystyle{IEEEtran}
\bibliography{ref}

@misc{nazari2024fortune,
  author       = {Samira Nazari and M. Taheri and
                  Ali Azarpeyvand and Mohsen Afsharchi and
                  Tara Ghasempouri and Christian Herglotz and
                  Masoud Daneshtalab and Maksim Jenihhin},
  title        = {{FORTUNE}: A Negative Memory Overhead
                  Hardware-Agnostic Fault Tolerance Technique
                  in {DNNs}},
  howpublished = {Authorea Preprints},
  year         = {2024},
}

@inproceedings{kodamanchili2025adaptive,
  author    = {Rama Mounika Kodamanchili and Natalia Cherezova
               and M. Taheri and Maksim Jenihhin},
  title     = {Adaptive Fault Resilience for Early-Exit {DNN}s},
  booktitle = {2025 IEEE International Test Conference in Asia
               (ITC-Asia)},
  year      = {2025},
  publisher = {IEEE},
}

@inproceedings{dosovitskiy2021vit,
  author    = {Alexey Dosovitskiy and Lucas Beyer and
               Alexander Kolesnikov and Dirk Weissenborn and
               Xiaohua Zhai and Thomas Unterthiner and
               Mostafa Dehghani and Matthias Minderer and
               Georg Heigold and Sylvain Gelly and
               Jakob Uszkoreit and Neil Houlsby},
  title     = {An Image is Worth 16$\times$16 Words:
               Transformers for Image Recognition at Scale},
  booktitle = {International Conference on Learning
               Representations (ICLR)},
  year      = {2021},
  url       = {https://arxiv.org/abs/2010.11929},
}

@inproceedings{vaswani2017attention,
  author    = {Ashish Vaswani and Noam Shazeer and Niki Parmar
               and Jakob Uszkoreit and Llion Jones and
               Aidan N. Gomez and {\L}ukasz Kaiser and
               Illia Polosukhin},
  title     = {Attention Is All You Need},
  booktitle = {Advances in Neural Information Processing
               Systems (NeurIPS)},
  volume    = {30},
  year      = {2017},
}

@article{helber2019eurosat,
  author  = {Patrick Helber and Benjamin Bischke and
             Andreas Dengel and Damian Borth},
  title   = {{EuroSAT}: A Novel Dataset and Deep Learning
             Benchmark for Land Use and Land Cover
             Classification},
  journal = {IEEE Journal of Selected Topics in Applied
             Earth Observations and Remote Sensing},
  volume  = {12},
  number  = {7},
  pages   = {2217--2226},
  year    = {2019},
  doi     = {10.1109/JSTARS.2019.2918242},
}

@article{furano2020edge,
  author  = {Gianluca Furano and Gabriele Meoni and
             Aubrey Dunne and David Moloney and
             V. Ferlet-Cavrois and
             Antonis Tavoularis and J. Byrne and
             L. Buckley and Mihalis Psarakis and
             Kay-Obbe Voss and Luca Fanucci},
  title   = {Towards the Use of Artificial Intelligence on
             the Edge in Space Systems: Challenges and
             Opportunities},
  journal = {IEEE Aerospace and Electronic Systems Magazine},
  volume  = {35},
  number  = {12},
  pages   = {44--56},
  year    = {2020},
  doi     = {10.1109/MAES.2020.3008278},
}

@inproceedings{denby2020orbital,
  author    = {Bradley Denby and Brandon Lucia},
  title     = {Orbital Edge Computing: Nanosatellite
               Constellations as a New Class of Computer
               System},
  booktitle = {Proceedings of the 25th International
               Conference on Architectural Support for
               Programming Languages and Operating Systems
               (ASPLOS)},
  pages     = {939--954},
  year      = {2020},
  doi       = {10.1145/3373376.3378473},
}

@article{mateo2023orbit,
  title={In-orbit demonstration of a re-trainable machine learning payload for processing optical imagery},
  author={Mateo-Garcia, Gonzalo and Veitch-Michaelis, Josh and Purcell, Cormac and Longepe, Nicolas and Reid, Simon and Anlind, Alice and Bruhn, Fredrik and Parr, James and Mathieu, Pierre Philippe},
  journal={Scientific Reports},
  volume={13},
  number={1},
  pages={10391},
  year={2023},
  publisher={Nature Publishing Group UK London}
}

@article{diana2024review,
  author    = {L. Diana and P. Dini},
  title     = {Review on Hardware Devices and Software
               Techniques Enabling Neural Network Inference
               Onboard Satellites},
  journal   = {Remote Sensing},
  volume    = {16},
  number    = {21},
  pages     = {3957},
  year      = {2024},
  publisher = {MDPI},
}

@article{shao2024spaceborne,
  author={Shao, Yingzhao and Wang, Junyi and Han, Xiaodong and Li, Yunsong and Li, Yaolin and Tao, Zhanpeng},
  title     = {Research on Spaceborne Neural Network
               Accelerator and Its Fault Tolerance Design},
  journal   = {Remote Sensing},
  volume    = {17},
  number    = {1},
  pages     = {69},
  year      = {2024},
  publisher = {MDPI},
}

@misc{tedeschi2025safeneureka,
  author       = {Riccardo Tedeschi and Luigi Ghionda and
                  Alessandro Nadalini and Yvan Tortorella and
                  Arpan Suravi Prasad and Luca Benini and
                  Davide Rossi and Francesco Conti},
  title        = {{Safe-NEureka}: A Hybrid Modular Redundant
                  {DNN} Accelerator for On-board Satellite
                  {AI} Processing},
  howpublished = {arXiv preprint arXiv:2602.04803},
  year         = {2026},
}

@article{baumann2005radiation,
  author  = {Robert C. Baumann},
  title   = {Radiation-Induced Soft Errors in Advanced
             Semiconductor Technologies},
  journal = {IEEE Transactions on Device and Materials
             Reliability},
  volume  = {5},
  number  = {3},
  pages   = {305--316},
  year    = {2005},
}

@inproceedings{reagen2018ares,
  author    = {Brandon Reagen and Udit Gupta and
               Lillian Pentecost and Paul Whatmough and
               Sae Kyu Lee and Niamh Mulholland and
               David Brooks and Gu-Yeon Wei},
  title     = {{ARES}: A Framework for Quantifying the
               Resilience of Deep Neural Networks},
  booktitle = {ACM/IEEE Design Automation Conference (DAC)},
  pages     = {1--6},
  year      = {2018},
}

@inproceedings{leveugle2009statistical,
  author    = {Regis Leveugle and Alexandre Calvez and
               Paolo Maistri and Pierre Vanhauwaert},
  title     = {Statistical Fault Injection: Quantified Error
               and Confidence},
  booktitle = {Design, Automation and Test in Europe
               Conference (DATE)},
  pages     = {502--506},
  year      = {2009},
  doi       = {10.1145/1874643.1874743},
}

@article{ruospo2025quantitative,
  author={Ruospo, Annachiara and Reorda, Matteo Sonza and Mariani, Riccardo and Sanchez, Ernesto},
  title   = {An Effective Iterative Statistical Fault
             Injection Methodology for Deep Neural Networks},
  journal = {IEEE Transactions on Computers},
  volume  = {74},
  pages   = {2431--2444},
  year    = {2025},
}

@inproceedings{wang2021tent,
  author    = {Dequan Wang and Evan Shelhamer and
               Shaoteng Liu and Bruno Olshausen and
               Trevor Darrell},
  title     = {Tent: Fully Test-Time Adaptation by Entropy
               Minimization},
  booktitle = {International Conference on Learning
               Representations (ICLR)},
  year      = {2021},
}

@misc{daniels2023efficient,
  author       = {Zachary A. Daniels and Jun Hu and
                  Michael Lomnitz and Phil Miller and
                  Aswin Raghavan and Joe Zhang and
                  Michael Piacentino and David Zhang},
  title        = {Efficient Model Adaptation for Continual
                  Learning at the Edge},
  howpublished = {arXiv preprint arXiv:2308.02084},
  year         = {2023},
}

@misc{christophides2024stochastic,
  author       = {Theodoros Christophides and Kyriakos Tolias
                  and Sotirios Chatzis},
  title        = {Continual Deep Learning on the Edge via
                  Stochastic Local Competition among
                  Subnetworks},
  howpublished = {arXiv preprint arXiv:2407.10758},
  year         = {2024},
}

@inproceedings{tung2026anatomy,
  author    = {Chung-Hsuan Tung and Yanxiang Huang and
               Nirmal Saxena and Philip Shirvani and
               Saurabh Hukerikar and Twinkle Jain and
               Abhishek Tyagi and Sanjay Gongalore},
  title     = {The Anatomy of Silent Data Corruption:
               {GPU} Error Pattern Study and Modeling
               Guidance},
  booktitle = {IEEE/IFIP International Conference on
               Dependable Systems and Networks (DSN),
               Industry Track},
  journal={arXiv preprint arXiv:2605.04213},
  year      = {2026},
}

@article{xue2023softerror,
  author={Xue, Xinghua and Liu, Cheng and Wang, Ying and Yang, Bing and Luo, Tao and Zhang, Lei and Li, Huawei and Li, Xiaowei},
  title   = {Soft Error Reliability Analysis of Vision
             Transformers},
  journal = {IEEE Transactions on Very Large Scale
             Integration (VLSI) Systems},
  volume  = {31},
  number  = {12},
  pages   = {2126--2136},
  year    = {2023},
}

@article{he2025fine,
  author={He, Jiajun and Liu, Yi and Xu, Changqing and Liao, Xinfang and Yang, Yintang},
  title   = {Fine-Grained Fault Sensitivity Analysis of
             Vision Transformers Under Soft Errors},
  journal = {Electronics},
  volume  = {14},
  pages   = {2418},
  year    = {2025},
}

@inproceedings{liao2025analyzing,
  author    = {En-Yu Liao and Ting-Chi Wang},
  title     = {Analyzing and Enhancing the Reliability of
               Vision Transformer Models Against Soft Errors},
  booktitle = {IEEE International Symposium on Circuits and
               Systems (ISCAS)},
  pages     = {1--5},
  year      = {2025},
}

@article{kirkpatrick2017overcoming,
  author  = {James Kirkpatrick and Razvan Pascanu and
             Neil Rabinowitz and Joel Veness and
             Guillaume Desjardins and Andrei A. Rusu and
             Kieran Milan and John Quan and Tiago Ramalho and
             Agnieszka Grabska-Barwinska and Demis Hassabis and
             Claudia Clopath and Dharshan Kumaran and Raia Hadsell},
  title   = {Overcoming Catastrophic Forgetting in Neural Networks},
  journal = {Proceedings of the National Academy of Sciences},
  volume  = {114},
  number  = {13},
  pages   = {3521--3526},
  year    = {2017},
  doi     = {10.1073/pnas.1611835114},
}

@article{li2018lwf,
  author  = {Zhizhong Li and Derek Hoiem},
  title   = {Learning without Forgetting},
  journal = {IEEE Transactions on Pattern Analysis and
             Machine Intelligence},
  volume  = {40},
  number  = {12},
  pages   = {2935--2947},
  year    = {2018},
  doi     = {10.1109/TPAMI.2017.2773081},
}

@misc{ba2016layernorm,
  author       = {Jimmy Lei Ba and Jamie Ryan Kiros and
                  Geoffrey E. Hinton},
  title        = {Layer Normalization},
  howpublished = {arXiv preprint arXiv:1607.06450},
  year         = {2016},
  url          = {https://arxiv.org/abs/1607.06450},
}

@inproceedings{zenke2017synaptic,
  author    = {Friedemann Zenke and Ben Poole and
               Surya Ganguli},
  title     = {Continual Learning Through Synaptic Intelligence},
  booktitle = {Proceedings of the 34th International
               Conference on Machine Learning (ICML)},
  pages     = {3987--3995},
  year      = {2017},
  url       = {https://arxiv.org/abs/1703.04200},
}

@inproceedings{rebuffi2017icarl,
  author    = {Sylvestre-Alvise Rebuffi and Alexander Kolesnikov
               and Georg Sperl and Christoph H. Lampert},
  title     = {{iCaRL}: Incremental Classifier and Representation
               Learning},
  booktitle = {IEEE/CVF Conference on Computer Vision and Pattern
               Recognition (CVPR)},
  pages     = {2001--2010},
  year      = {2017},
}

@article{friasdominguez2026dependability,
  author  = {L. Frias-Dom{\'i}nguez and J. M. Badia and
             G. Le{\'o}n and others},
  title   = {Dependability Analysis and Hardening of Vision
             Transformers Against Soft Errors},
  journal = {The Journal of Supercomputing},
  volume  = {82},
  pages   = {331},
  year    = {2026},
  doi     = {10.1007/s11227-026-08373-0},
}

@article{ridnik2021,
  title={Imagenet-21k pretraining for the masses},
  author={Ridnik, Tal and Ben-Baruch, Emanuel and Noy, Asaf and Zelnik-Manor, Lihi},
  journal={arXiv preprint arXiv:2104.10972},
  year={2021}
}

@article{SENTRY,
  title={SENTRY: Statistical Reliability Analysis of Vision Transformers Under Soft Errors},
  author={Bhaduri, Pramit Kumar and Taheri, Mahdi and Nazari, Samira and Jenihhin, Maksim and Herglotz, Christian and Hubner, Michael},
  journal={arXiv preprint arXiv:2606.07620},
  year={2026}
}

@article{wang2024comprehensive,
  title={A comprehensive survey of continual learning: Theory, method and application},
  author={Wang, Liyuan and Zhang, Xingxing and Su, Hang and Zhu, Jun},
  journal={IEEE transactions on pattern analysis and machine intelligence},
  volume={46},
  number={8},
  pages={5362--5383},
  year={2024},
  publisher={IEEE}
}

@inproceedings{liu2025enabling,
  title={Enabling real-time inference in online continual learning via device-cloud collaboration},
  author={Liu, Haibo and Gong, Chen and Zheng, Zhenzhe and Liu, Shengzhong and Wu, Fan},
  booktitle={Proceedings of the ACM on Web Conference 2025},
  pages={2043--2052},
  year={2025}
}

@article{sze2017efficient,
  title={Efficient processing of deep neural networks: A tutorial and survey},
  author={Sze, Vivienne and Chen, Yu-Hsin and Yang, Tien-Ju and Emer, Joel S},
  journal={Proceedings of the IEEE},
  volume={105},
  number={12},
  pages={2295--2329},
  year={2017},
  publisher={Ieee}
}

@article{kadi2018general,
  title={General-purpose computing with soft GPUs on FPGAs},
  author={Kadi, Muhammed Al and Janssen, Benedikt and Yudi, Jones and Huebner, Michael},
  journal={ACM Transactions on Reconfigurable Technology and Systems (TRETS)},
  volume={11},
  number={1},
  pages={1--22},
  year={2018},
  publisher={ACM New York, NY, USA}
}

@inproceedings{21,
  title={HMR-NEureka: Hybrid modular redundancy DNN acceleration in heterogeneous RISC-V SoCs},
  author={Ghionda, Luigi and Tedeschi, Riccardo and Tortorella, Yvan and Prasad, Arpan Suravi and Rossi, Davide and Benini, Luca and Conti, Francesco},
  booktitle={2025 IEEE Computer Society Annual Symposium on VLSI (ISVLSI)},
  volume={1},
  pages={1--6},
  year={2025},
  organization={IEEE}
}

@article{22,
  title={Safe-NEureka: a Hybrid Modular Redundant DNN Accelerator for On-board Satellite AI Processing},
  author={Tedeschi, Riccardo and Ghionda, Luigi and Nadalini, Alessandro and Tortorella, Yvan and Prasad, Arpan Suravi and Benini, Luca and Rossi, Davide and Conti, Francesco},
  journal={arXiv preprint arXiv:2602.04803},
  year={2026}
}

@inproceedings{17,
  title={FlexGrip: A soft GPGPU for FPGAs},
  author={Andryc, Kevin and Merchant, Murtaza and Tessier, Russell},
  booktitle={2013 International Conference on Field-Programmable Technology (FPT)},
  pages={230--237},
  year={2013},
  organization={IEEE}
}

@article{7,
  title={Radiation-induced soft errors in advanced semiconductor technologies},
  author={Baumann, Robert C},
  journal={IEEE Transactions on Device and materials reliability},
  volume={5},
  number={3},
  pages={305--316},
  year={2005},
  publisher={IEEE}
}

\end{document}